\documentclass[12pt,twocolumn,preprint]{aastex}
\usepackage{fullpage}
\usepackage{graphicx}
\usepackage{ulem}
\def\sun{\odot}
\def\earth{\oplus}
\input epsf.tex
\begin{document}

 \title{No Planet Left Behind: Investigating Planetary Diversity and
Architecture with SIM Lite}
 \author{
 S. R. Kulkarni\altaffilmark{1}, 
H. E. Schlichting\altaffilmark{1},B. M. S. Hansen\altaffilmark{2},
 \&\
J. Catanzarite\altaffilmark{3}
 }
 \altaffiltext{1}{Department of Astronomy,
California Institute of Technology, Pasadena, CA 91125}
\altaffiltext{2}{Department of Physics \& Astronomy, University of California
Los Angeles, Los Angeles, CA 90095}
 \altaffiltext{3}{Jet Propulsion
Laboratory, Pasadena, CA 91109}

\section{Abstract}
The evidence is mounting that star formation necessarily involves planet
formation. We clearly have a vested interest in finding other Earths but a
true understanding of planet formation requires completing the census and
mapping planetary architecture in all its grandeur and diversity. Here, we
show that a 2000-star survey undertaken with SIM Lite will uniquely probe
planets around B-A-F stars, bright and binary stars and white dwarfs. In
addition, we show that the high precision of SIM Lite allows us to gain unique
insights into planet formation via accurate measurements of mutual
inclinations.


\section{Introduction}
Our understanding of extrasolar planetary systems has grown
exponentially over the past decade and half.
In addition to familiar designations of rocky planets,
giant planets and icy giants we now have 
new names such as ``Hot Jupiters'', ``Eccentric Giants'',
``Hot Neptunes'' and ``Super Earths''. 

The first wave of these discoveries was driven by precision
Radial Velocity (RV) studies. The transit method is now contributing
handsomely to the detailed studies (radius, composition) of the
hot Jupiters. 
COROT and Kepler (launch in 3 weeks!) will determine
the statistics of rocky planets.

Recently,  the ExoPlanet Task Force (ExoPTF)\footnote{\texttt{http://www.nsf.gov/mps/ast/exoptf.jsp}}
reviewed the state of the field. Their strategy consisted of addressing the following fundamental questions (in priority order) over the next decade and half:
\begin{enumerate}
\item What are the physical characteristics of planets in the habitable zones
around bright, nearby stars?
\item What is the architecture of planetary systems?
\item When, how and in what environments are planets formed?
\end{enumerate}
Other white papers (e.g. Marcy-Shao, Traub-Kastings, Beichman) 
address the first and last question. 
Here, we address the second question.

\section{Planetary Diversity \&\ Architecture}
For the Solar system, the
observations and
measurements strongly support the bottom-up
(dust to rocks to planetary cores), also known as
Safronov model for planet formation. In contrast, the prevailing
hypothesis for the formation of brown dwarfs (and stars) is a
top-down (gravitational condensation) scenario.

The discovery of 51 Pegasi b, a Jupiter with an orbital separation
of only 0.05\,AU (as opposed to 5.2\,AU for Jupiter), was
a dramatic illustration of the limitations of the standard model for planet
formation.

Observations have now established a
strong correlation
between the metallicity of stars and the occurrence of an planet
(identified by RV approach). The sense of connection 
(metals to planets) as well as whether this correlation is proportional 
(low metallicity, fewer or lower mass planets as opposed to a sharp transition) 
are being debated heavily.

It is well known that most stars are in binary or multiple
systems. A full understanding of planet formation should naturally
address the issue of planets around and in binary (and
multiple) star systems.

Finally, the current extra-solar planet sample is dominated by those
found using the RV technique, namely stars with spectral type FGK. 
OBA stars have no strong absorption features and M dwarfs
have prominent lines but primarily in the near-IR. Binaries with
small angular separation pose additional difficulties for observations.

These gaps in our knowledge show the importance of a comprehensive search
for planets in every conceivable ecological niche: stars with varying
metallicity, binary stars and stars across the entire mass spectrum.

Apart from these astrophysical ``biases'', the search techniques
have their own biases:
RV and transits favor close in planets whereas astrometry gains ascendency
with longer period planets. Both RV and astrometry are limited by
the duration of the survey. Micro-lensing, while sensitive, is limited
to statistical studies. Imaging techniques will be valuable but 
the meaningfully powerful instruments are a decade away.

Mapping planetary architecture would be immensely aided by
having sensitive astrometric measurements.
Fortunately, recent advances
in technology will soon see astrometry fulfilling its expected
promise. 

\begin{figure}[htbp] 
   \centering
   \includegraphics[width=3.3in]{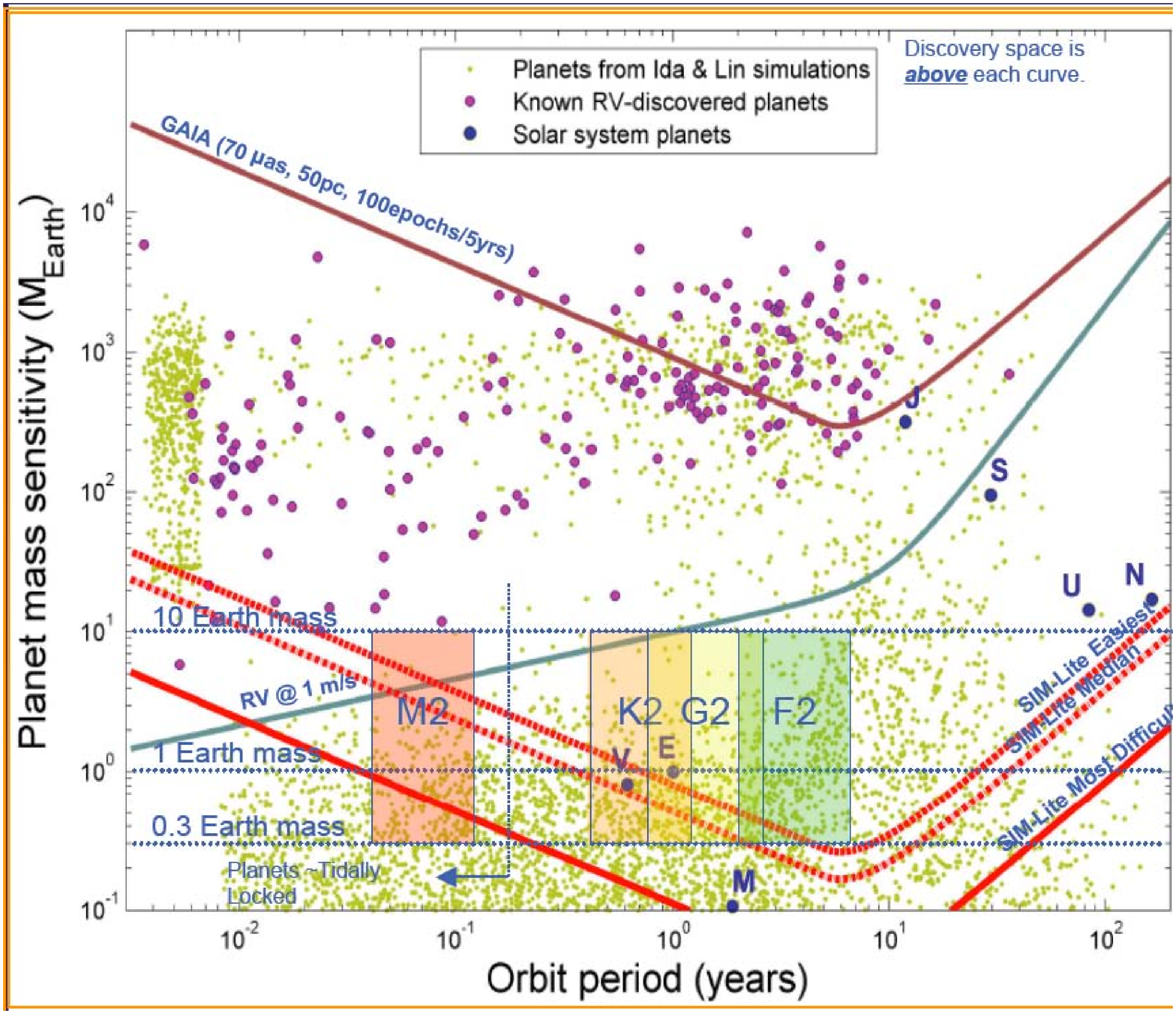}
   \caption{\small Phase Space of SIM Habitable Zone planet search
and the GAIA planet search.}
   \label{fig:SIMGAIA}
\end{figure}

\section{The Decade of Astrometry}
Ground-based interferometers have already demonstrated 
GAIA-like single-epoch (or better)  performance for close
binaries ($20\,\mu$arcsec; \citet{mlf+08}); seeing-limited
imaging and HST/FGS observations have achieved precision of sub-milliarcsecond 
for relative astrometry \citep{l06,psw+06,bmb08};
and adaptive optics observations show great promise of
beating $100\,\mu$arcsec \citep{cbk08}.

The main limitation for ground-based interferometric and AO astrometry
is the availability of suitably bright reference stars.
As a result 
ground-based interferometry is ideally suited to exploring planets
in binary systems. AO observations with large telescopes is well suited
to probing planets around faint targets especially at low
Galactic latitudes (M dwarfs, brown dwarfs). 
But for most stars, the requirement of reference stars 
makes space based astrometry a must. This basic conclusion has
been discussed and reaffirmed by two decadal reviews (1990 and 2000) and
again reaffirmed recently by ExoPTF.

GAIA (expected launch of late 2011) is expected to achieve single epoch
astrometric precision of $55\,\mu$arcsec (for the range 6--13\,mag).  With an
average of 84 visits to an object GAIA has very good sensitivity to detect
Jupiter mass objects around a very large number stars.

SIM Lite is designed for both wide and narrow angle astrometry.
Three planet searches have been envisaged with SIM Lite: an ultra-deep
sub-microarcsecond search of nearby Earth-like planets around nearby
Sun-like stars (PI: Shao, PI: Marcy; hereafter, HZ search), 
a search for planets around young stars (PI: Beichman) and a
broad search. This latter search is the topic of this paper.
The phase space covered by GAIA and the SIM habitable zone search is shown
in Figure~\ref{fig:SIMGAIA}. 

\begin{figure}[htbp] 
   \centering
   \includegraphics[width=3.5in]{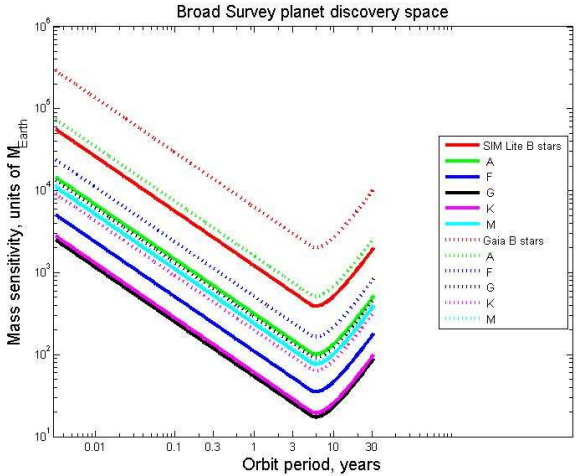} 
   \caption{\small Phase space of SIM Lite Broad Survey with High Precision relative to that
   of GAIA. SIM Lite enjoys a clear advantage over GAIA for BAF stars. Nearby
GK and some F and M stars will be observed intensively by the SIM Lite
HZ search (see previous Figure).}
   \label{fig:BSHP}
\end{figure}

Over the range 0--13\,mag SIM Lite can easily achieve 5\,$\mu$arcsec
single-epoch precision. With 10\% of SIM Lite time one can survey nearly 2,000
stars at this single-epoch sensitivity (visiting each star 150 times).  We
call this as the ``Broad Survey with High Precision'' (BSHP for short) and
discuss the potential astrophysical returns of this survey.  The relative
phase space between GAIA and BSHP is shown in Figure~\ref{fig:BSHP}.

\section{Planets around B- and A-type Stars} 
RV studies, by necessity, have targeted FGK stars.  For example,
the bulk of the California and Carnegie Planet Search probe the
mass range 0.8--1.2\,$M_\odot$ \citep{vf05,tfs+07}.  The intermediate-
and high-mass main sequence stars ($M_*>1.4\,M_\odot$) suffer from
fewer spectral lines, rapid rotation and surface inhomogeneities
\citep{sbm+98,nmm+03,glu+05,w05}. By cleverly observing
evolved versions of these stars, \citet{jbm+07} find that the planet
occurrence rate increases with increasing stellar mass.

SIM Lite is well positioned to undertake a comprehensive survey
of hundreds of type A and B stars.
For example, SIM Lite will be able to detect a
$19 M_{\earth}$ planet on a 4 year orbit around a $2 M_{\sun}$ A-type star
located at 30\,pc with 150 50-second visits. Similarly, a $130 M_{\earth}$
planet can be detected around a $6 M_{\sun}$ B-type star located at 100pc. 

\begin{figure}[ht]
   \centering
   \includegraphics[width=3in]{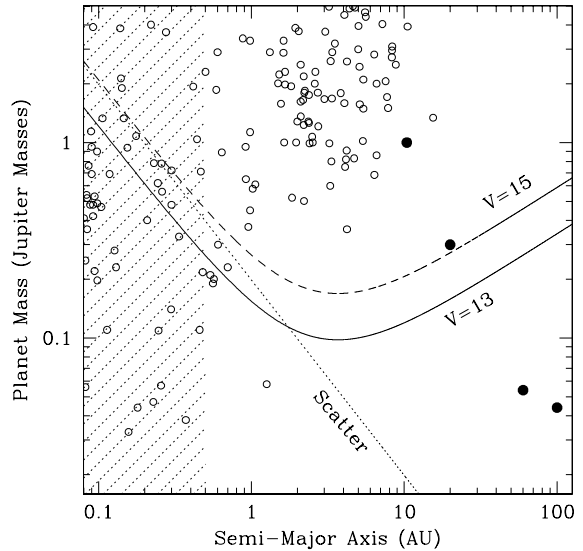} 
\caption{\small The discovery space for a five-year astrometric planet search
around white dwarfs (distance 10 pc).
See text (\S\ref{sec:DAZd}) for explanation of the dotted line.
The solid points indicate
the positions of the solar system planets assuming the Sun loses
half its mass.
Similarly, the open circles indicate the positions of the detected
radial velocity planets\citep{bwm+} if they  spiraled out by a
factor of two during the evolution to a white dwarf. 
Planets in the shaded region will be swallowed up by the red giant
precursor phase.
\label{fig:Acc}} 
\end{figure}

\section{Planets and Host Star Metalicity}
Jovian-like planets are preferentially found around metal rich stars
\citep{sim+04,fv05}. Recent observational findings suggest that this
well-established result does not hold for Neptunian-like
planets. \citet{ssm+08} find a wide spread in metallicities for stars hosting
Neptunian-like planets and find that the Jupiter-to-Neptune ratio is higher
for higher metalicity stars. These results suggest that the mass of the
largest planet in any given system is determined by the metalicity of the host
star. This trend is expected from planet formation theories based on the
core-accretion process provided that the host star metalicity is
representative of the metalicity in the planetesimal disk \citep{bbe+06}. This
suggests that lower-mass planets might even be preferentially found orbiting
metal-poor stars.  A SIM Lite survey (discussed in the previous section) will
be able to test whether the planet-metalicity relation holds for A- and B-type
stars.

\section{Binary \&\ Bright Stars}
GAIA is not able to observe stars brighter than 6 mag. For stars with
$6<V<13$ saturation is avoided by dumping the accumulated charge.
As a result GAIA has a flat astrometric performance to 13 mag (after
which photon noise becomes important). For bright stars a surrounding region
(proportional to the brightness) is not observable. This limitation
means that a range of binary stars (with at least one bright companion)
are not accessible to GAIA.
 
The absolute V magnitudes of dwarfs is as follows: G5 (5.1), G0 (4.4),
F5 (3.4), F0 (2.6), A5 (2.0), A0 (0.7), B5 (-1.1) and B0 (-4.7).
The following stars are not accessible to GAIA: $\alpha$ Centauri
(G2V), Sirius~A (sp type A0),
Altair (A7),  Procyon (F5), Regulus (B8),
Alkaid (B3) 
and so on.

Next, the neighborhood restriction discussed above excludes
planet searches around fainter members of a bright star. Ground-based interferometers equipped with
dual-beam correlators (phase referencing) have already demonstrated
GAIA-like precision for planet searches (cf. the PHASES project,
\cite{mlf+08}). A SIM Lite + VLTI program targeting suitable binaries
would offer the best of both worlds:
high precision over a 5-yr period and a 25-yr search for distant
companions.

\section{Planets around White Dwarfs}
Eventually, the majority of stars evolve to become white dwarfs.
It is a natural question to ask what happens to a pre-existing
planetary system as the star evolves. For planets sufficiently far
from their parent stars, the mass loss in the later evolutionary
phases results in an adiabatic expansion of the orbit, so that the
planets spiral outwards, but otherwise remain bound. For the closest
planets, however, the out-spiral is not sufficiently fast, and the
planet is, at some point engulfed by the expanding host.  Tidal
interactions between the star and the planet also influence where
this boundary lies. In addition to the general astrophysical interest,
this question has an anthropocentric (if morbid) interest, in that
studies show that the long-term survival of Earth in the face of
the Sun's evolution is uncertain, as it lies near the boundary,
where different treatments of tidal and wind effects can yield
different answers\citep{rtl+}; see also 
\citep{dl,vl}.

Lack of strong and/or narrow absorption lines limit RV precision to 10
km\,s$^{-1}$ (except for the very special cases of pulsating white dwarfs
\citep{mwd+}). Furthermore, the red-giant phase of the host star leads to
spiraling of inner planets. Astrometry is ideally suited to probe planets
around white dwarfs.  The astrometric method is further favored by the
proximity of white dwarfs (122 within the local 20\,pc; \cite{hso+08}).

A five year astrometric program\footnote{Two hundred visits
with SIM Lite over a five year period. Integration time 
of 15s (V=13; solid line) and 30-s (V=15; dashed line);
see Figure 3}.
probes precisely the original $\sim
1 AU$ region of anthropocentric interest. Assuming a traditional
initial-final mass relation $M_{f} = 0.49 M_{\odot} \exp (0.095
M_{i})$ (e.g.\citet{w}), conservation of angular momentum during
the main sequence to white dwarf transition implies that a final
(circular) orbit with a period of five years around the white dwarf
corresponds to an original semi-major axis
\begin{equation} 
	a_i = 1.05 AU
	\frac{e^{0.127 M_i}}{M_i} \left( \frac{P_f}{5{\rm\,yr}} \right)^{2/3}.
\end{equation}
A SIM Lite white dwarf planet search will probe planets down to 
roughly a Neptune mass at original distances
0.5--2 AU (Figure~\ref{fig:Acc}).

\subsection{DAZd White Dwarfs}
\label{sec:DAZd}

Approximately 2\% of all white dwarfs with cooling ages $< 0.5$~Gyr show
evidence for an infrared excess \citep{fjz} and some show evidence for metal
pollution \citep{kvl+,jfz}. These are attributed to the tidal disruption of a
planetary minor body, either a comet or asteroid \citep{afs,j} to form a disk
that reprocesses stellar light and slowly accretes onto the star.

Because a white dwarf progenitor swells to radii $\sim 1 AU$ during
prior evolutionary stages, asteroids that approach close enough to
be tidally disrupted must be scattered inwards at late times by
planetary bodies \citep{ds}. Planets large enough to scatter
significantly without accreting must have a mass:
$M > 20 M_{\oplus} \left( \frac{a}{1 AU} \right)^{-1}$ where $a$ is the   
semi-major axis (shown as dotted line in Figure~\ref{fig:Acc}).
The SIM Lite white dwarf program
will probe a significant
fraction of the parameter space occupied by planets that 
generate these dusty disks through asteroid scattering. 
The sample of $V<15$ white dwarfs is large enough  to test 
the hypothesis that most of this particular class of white dwarfs
have surviving planetary systems.

\section{Insight through Precision}
It has long been appreciated that mutual inclination (the inclination
of planets with respect to each other) and eccentricity give fundamental
insight into details of planet formation.  Astrometry (and imaging)
is uniquely suited to measuring inclinations.

Among the great variety of planetary systems uncovered by the radial velocity
studies are a number of multiple planet systems (32 as of Feb 14, 2009). 
Sometimes interactions result in resonant states. For example,
a 3:1 mean motion resonance is claimed in HD~60532 \citep{dlg+,lc}.

SIM Lite is particularly well suited to probing these subtle but key diagnostic
dynamical clues for planets with $a>0.5\,$AU. True mass determination is
clearly essential for a correct understanding of the dynamics of the system
(stability, identification of mean motion resonances and secular resonances).
Next, the mutual inclinations of eccentric planets shed light on the prior
evolution of the system (e.g. diffusive scattering processes should lead to
approximate energy equipartition in radial and vertical motions, whereas
resonant processes need not do so). In addition, determining the mass ratio
and resonance configuration of multiple planet systems will place constraints
on the strength of eccentricity damping during migration and the rate of
planetary the migration itself \citep{lt04}.

Separately, should the orbit of a planet be inclined significantly with
respect to the binary orbit, Kozai oscillations can significant affect the
orbital parameters of the system \citep{htt,wm}. Furthermore, a statistically
significant correlation between the sense of rotation for stellar orbits and
planetary orbits may provide information on the degree to which the binarity
affects the formation of a planetary system.


\begin{thebibliography}{34}
\expandafter\ifx\csname natexlab\endcsname\relax\def\natexlab#1{#1}\fi
\small
\baselineskip=-1pt
\parskip=-1pt

\bibitem[{{Alcock} {et~al.}(1986){Alcock}, {Fristrom}, \& {Siegelman}}]{afs}
{Alcock}, C., {Fristrom}, C.~C., \& {Siegelman}, R. 1986, \apj, 302, 462

\bibitem[{{Bean} {et~al.}(2006){Bean}, {Benedict}, \& {Endl}}]{bbe+06}
{Bean}, J.~L., {Benedict}, G.~F., \& {Endl}, M. 2006, \apjl, 653, L65

\bibitem[{{Benedict} {et~al.}(2008){Benedict}, {McArthur}, \& {Bean}}]{bmb08}
{Benedict}, G.~F., {McArthur}, B.~E., \& {Bean}, J.~L. 2008, in IAU Symposium,
  Vol. 248, IAU Symposium, 23--29

\bibitem[{{Butler} {et~al.}(2006){Butler}, {Wright}, {Marcy}, {Fischer},
  {Vogt}, {Tinney}, {Jones}, {Carter}, {Johnson}, {McCarthy}, \&
  {Penny}}]{bwm+}
{Butler}, R.~P., {Wright}, J.~T., {Marcy}, G.~W., {Fischer}, D.~A., {Vogt},
  S.~S., {Tinney}, C.~G., {Jones}, H.~R.~A., {Carter}, B.~D., {Johnson}, J.~A.,
  {McCarthy}, C., \& {Penny}, A.~J. 2006, \apj, 646, 505

\bibitem[{{Cameron} {et~al.}(2009){Cameron}, {Britton}, \& {Kulkarni}}]{cbk08}
{Cameron}, P.~B., {Britton}, M.~C., \& {Kulkarni}, S.~R. 2009, \aj, 137, 83

\bibitem[{{Debes} \& {Sigurdsson}(2002)}]{ds}
{Debes}, J.~H. \& {Sigurdsson}, S. 2002, \apj, 572, 556

\bibitem[{{Desort} {et~al.}(2008){Desort}, {Lagrange}, {Galland}, {Beust},
  {Udry}, {Mayor}, \& {Lo Curto}}]{dlg+}
{Desort}, M., {Lagrange}, A.-M., {Galland}, F., {Beust}, H., {Udry}, S.,
  {Mayor}, M., \& {Lo Curto}, G. 2008, \aap, 491, 883

\bibitem[{{do Nascimento} {et~al.}(2003){do Nascimento}, {Canto Martins},
  {Melo}, {Porto de Mello}, \& {De Medeiros}}]{nmm+03}
{do Nascimento}, Jr., J.~D., {Canto Martins}, B.~L., {Melo}, C.~H.~F., {Porto
  de Mello}, G., \& {De Medeiros}, J.~R. 2003, \aap, 405, 723

\bibitem[{{Duncan} \& {Lissauer}(1998)}]{dl}
{Duncan}, M.~J. \& {Lissauer}, J.~J. 1998, Icarus, 134, 303

\bibitem[{{Farihi} {et~al.}(2009){Farihi}, {Jura}, \& {Zuckerman}}]{fjz}
{Farihi}, J., {Jura}, M., \& {Zuckerman}, B. 2009, ArXiv e-prints

\bibitem[{{Fischer} \& {Valenti}(2005)}]{fv05}
{Fischer}, D.~A. \& {Valenti}, J. 2005, \apj, 622, 1102

\bibitem[{{Galland} {et~al.}(2005){Galland}, {Lagrange}, {Udry}, {Chelli},
  {Pepe}, {Queloz}, {Beuzit}, \& {Mayor}}]{glu+05}
{Galland}, F., {Lagrange}, A.-M., {Udry}, S., {Chelli}, A., {Pepe}, F.,
  {Queloz}, D., {Beuzit}, J.-L., \& {Mayor}, M. 2005, \aap, 443, 337

\bibitem[{{Holberg} {et~al.}(2008){Holberg}, {Sion}, {Oswalt}, {McCook},
  {Foran}, \& {Subasavage}}]{hso+08}
{Holberg}, J.~B., {Sion}, E.~M., {Oswalt}, T., {McCook}, G.~P., {Foran}, S., \&
  {Subasavage}, J.~P. 2008, \aj, 135, 1225

\bibitem[{{Holman} {et~al.}(1997){Holman}, {Touma}, \& {Tremaine}}]{htt}
{Holman}, M., {Touma}, J., \& {Tremaine}, S. 1997, \nat, 386, 254

\bibitem[{{Johnson} {et~al.}(2007){Johnson}, {Butler}, {Marcy}, {Fischer},
  {Vogt}, {Wright}, \& {Peek}}]{jbm+07}
{Johnson}, J.~A., {Butler}, R.~P., {Marcy}, G.~W., {Fischer}, D.~A., {Vogt},
  S.~S., {Wright}, J.~T., \& {Peek}, K.~M.~G. 2007, \apj, 670, 833

\bibitem[{{Jura}(2003)}]{j}
{Jura}, M. 2003, \apjl, 584, L91

\bibitem[{{Jura} {et~al.}(2007){Jura}, {Farihi}, \& {Zuckerman}}]{jfz}
{Jura}, M., {Farihi}, J., \& {Zuckerman}, B. 2007, \apj, 663, 1285

\bibitem[{{Kilic} {et~al.}(2006){Kilic}, {von Hippel}, {Leggett}, \&
  {Winget}}]{kvl+}
{Kilic}, M., {von Hippel}, T., {Leggett}, S.~K., \& {Winget}, D.~E. 2006, \apj,
  646, 474

\bibitem[{{Laskar} \& {Correia}(2009)}]{lc}
{Laskar}, J. \& {Correia}, A.~C.~M. 2009, ArXiv e-prints

\bibitem[{{Lazorenko}(2006)}]{l06}
{Lazorenko}, P.~F. 2006, \aap, 449, 1271

\bibitem[{{Lee} \& {Thommes}(2004)}]{lt04}
{Lee}, M.~H. \& {Thommes}, E.~W. 2004, in Bulletin of the American Astronomical
  Society, Vol.~36, Bulletin of the American Astronomical Society, 1152--+

\bibitem[{{Mullally} {et~al.}(2008){Mullally}, {Winget}, {Degennaro},
  {Jeffery}, {Thompson}, {Chandler}, \& {Kepler}}]{mwd+}
{Mullally}, F., {Winget}, D.~E., {Degennaro}, S., {Jeffery}, E., {Thompson},
  S.~E., {Chandler}, D., \& {Kepler}, S.~O. 2008, \apj, 676, 573

\bibitem[{{Muterspaugh} {et~al.}(2008){Muterspaugh}, {Lane}, {Fekel},
  {Konacki}, {Burke}, {Kulkarni}, {Colavita}, {Shao}, \&
  {Wiktorowicz}}]{mlf+08}
{Muterspaugh}, M.~W., {Lane}, B.~F., {Fekel}, F.~C., {Konacki}, M., {Burke},
  B.~F., {Kulkarni}, S.~R., {Colavita}, M.~M., {Shao}, M., \& {Wiktorowicz},
  S.~J. 2008, \aj, 135, 766

\bibitem[{{Pravdo} {et~al.}(2006){Pravdo}, {Shaklan}, {Wiktorowicz},
  {Kulkarni}, {Lloyd}, {Martinache}, {Tuthill}, \& {Ireland}}]{psw+06}
{Pravdo}, S.~H., {Shaklan}, S.~B., {Wiktorowicz}, S.~J., {Kulkarni}, S.,
  {Lloyd}, J.~P., {Martinache}, F., {Tuthill}, P.~G., \& {Ireland}, M.~J. 2006,
  \apj, 649, 389

\bibitem[{{Rasio} {et~al.}(1996){Rasio}, {Tout}, {Lubow}, \& {Livio}}]{rtl+}
{Rasio}, F.~A., {Tout}, C.~A., {Lubow}, S.~H., \& {Livio}, M. 1996, \apj, 470,
  1187

\bibitem[{{Saar} {et~al.}(1998){Saar}, {Butler}, \& {Marcy}}]{sbm+98}
{Saar}, S.~H., {Butler}, R.~P., \& {Marcy}, G.~W. 1998, \apjl, 498, L153+

\bibitem[{{Santos} {et~al.}(2004){Santos}, {Israelian}, \& {Mayor}}]{sim+04}
{Santos}, N.~C., {Israelian}, G., \& {Mayor}, M. 2004, \aap, 415, 1153

\bibitem[{{Sousa} {et~al.}(2008){Sousa}, {Santos}, {Mayor}, {Udry},
  {Casagrande}, {Israelian}, {Pepe}, {Queloz}, \& {Monteiro}}]{ssm+08}
{Sousa}, S.~G., {Santos}, N.~C., {Mayor}, M., {Udry}, S., {Casagrande}, L.,
  {Israelian}, G., {Pepe}, F., {Queloz}, D., \& {Monteiro}, M.~J.~P.~F.~G.
  2008, \aap, 487, 373

\bibitem[{{Takeda} {et~al.}(2007){Takeda}, {Ford}, {Sills}, {Rasio}, {Fischer},
  \& {Valenti}}]{tfs+07}
{Takeda}, G., {Ford}, E.~B., {Sills}, A., {Rasio}, F.~A., {Fischer}, D.~A., \&
  {Valenti}, J.~A. 2007, \apjs, 168, 297

\bibitem[{{Valenti} \& {Fischer}(2005)}]{vf05}
{Valenti}, J.~A. \& {Fischer}, D.~A. 2005, \apjs, 159, 141

\bibitem[{{Villaver} \& {Livio}(2007)}]{vl}
{Villaver}, E. \& {Livio}, M. 2007, \apj, 661, 1192

\bibitem[{{Wood}(1992)}]{w}
{Wood}, M.~A. 1992, \apj, 386, 539

\bibitem[{{Wright}(2005)}]{w05}
{Wright}, J.~T. 2005, \pasp, 117, 657

\bibitem[{{Wu} \& {Murray}(2003)}]{wm}
{Wu}, Y. \& {Murray}, N. 2003, \apj, 589, 605

\end{thebibliography}
\end{document}